# Intrinsic bulk quantum oscillations in a bulk unconventional insulator SmB$_6$

M. Hartstein[1], H. Liu[1], Y.-T. Hsu[1], B. S. Tan[1],
M. Ciomaga Hatnean[2], G. Balakrishnan[2], & Suchitra E. Sebastian[1†]

[1]Cavendish Laboratory, Cambridge University, Cambridge CB3 0HE, UK,
[2]Department of Physics, University of Warwick, Coventry, CV4 7AL, UK.

July 2, 2020

**The finding of bulk quantum oscillations in the bulk Kondo insulator SmB$_6$ [1], which has been proposed to be a correlated topological insulator [2], proved a considerable surprise. The subsequent measurement of bulk quantum oscillations in other correlated insulators including YbB$_{12}$ [3, 4] have lent support to our discovery of a class of unconventional insulators that are host to bulk quantum oscillations, of which SmB$_6$ was the first example. Here we perform a series of experiments to examine evidence for the intrinsic character of bulk quantum oscillations in floating zone-grown single crystals of SmB$_6$ that have been the subject of our quantum oscillation studies thus far [1]. We present results of experiments including chemical composition analysis, magnetisation, thermal conductivity, electrical transport, and heat capacity on floating zone-grown single crystals of SmB$_6$, and a series of quantum oscillation experiments as a function of magnetic field, temperature, and magnetic field-orientation on single crystals of floating-zone grown**

SmB$_6$, LaB$_6$, and elemental Aluminium. Results of these experimental studies establish the intrinsic origin of quantum oscillations from the bulk of pristine floating zone-grown single crystals of SmB$_6$. The origin of the underlying bulk Fermi surface that bears close similarity with the unhybridised Fermi surface in metallic hexaborides [1, 5, 6] despite the bulk insulating character of SmB$_6$ is thus at the heart of a theoretical mystery.

# Results

## Pristine chemical composition of floating zone-grown single crystals of SmB$_6$ used for quantum oscillation measurements

Given the history of inclusions which have been shown to yield associated quantum oscillations in flux-grown samples of other systems (e.g. UBe$_{13}$ [7, 8], CaB$_6$ [9]), we chose to study quantum oscillations in floating zone-grown single crystals of SmB$_6$ which are not grown out of external flux in ref. [1] and in this study. Here we demonstrate the high quality of floating zone-grown single crystals of SmB$_6$ used for our quantum oscillation measurements, and place constraints on any impurity content using a combination of experimental techniques. To quantify the bulk elemental composition of our samples, we employ inductively coupled plasma optical emission spectrometry (ICP-OES) [10]. The bulk chemical composition analysis (shown in Table 1) indicates very high purity, with any extrinsic chemical content (including Aluminium and Gadolinium) other than Tellurium and Yttrium constrained below the level of detection ($\approx 0.01\%$at in most cases).

Suggestions of magnetic inclusions such as Gadolinium in single crystals of SmB$_6$ have been put forward based on studies of SmB$_6$ single crystals impregnated with Gadolinium impurities [12, 13, 14, 15, 16, 17, 18]. Here we present low temperature magnetisation measurements of our single crystals of floating zone-grown SmB$_6$ to place limits on any magnetic impurity content. For a paramagnet such as SmB$_6$, the magnetisation is expected to be linear as a function of an applied magnetic field. Any magnetic impurities in a paramagnet would be expected to yield a deviation from linearity, described by the Langevin function. The extent of deviation from linearity may be used to place an upper limit on any magnetic impurity content, which method has been used in the study in ref. [12]. Measurements on floating zone-grown single crystals of SmB$_6$ used in our quantum oscillation measurements yield a linear magnetisation [5], with a deviation from linearity consistent with a magnetic impurity concentration level

| Parts per million range | Elements |
|---|---|
| limit of detection (100) | Al, Au, As, Ba, Be, Bi, Ca, Cd, Ce, Co, Cr, Cu, Dy, Eu, Fe, Gd, Ge, Hf, Hg, Ho, Ir, K, La, Li, Lu, Mg, Mn, Mo, Na, Nb, Nd, Ni, Os, P, Pb, Pd, Pt, Rb, Re, Ru, S, Sb, Sc, Se, Si, Sn, Sr, Ta, Tb, Ti, Tl, Tm, U, V, Yb, Zn, Zr. |
| limit of detection (400) | Ag, Er, Ga, Nd, Se. |
| 100-1000 | Te, Y. |
| 1000-10000 | |
| 10000 or >1%at | B, Sm. |

Table 1: Limits on impurity concentration in parts per million in floating zone-grown $SmB_6$ single crystals used for our quantum oscillation measurements [11] as found by inductively coupled plasma optical emission spectrometry (ICP-OES). ICP-OES was performed by Exeter Analytical UK Ltd., where $\approx$ 50 mg of single crystals were digested in a nitric acid matrix using microwaves and introduced to the spectrometer with internal standards to aid precision. Extrinsic impurity content other than Tellurium and Yttrium is ruled out to within the detection limit ($\approx$ 0.01%at in most cases).

limited to within 0.04%at (Fig. 2(b)).

The high-temperature phonon peak marks the regime where the thermal conductivity is limited by impurities and defects, and the size of this peak provides a measure of the mean free path, thus indicating sample quality. Fig. 1 shows a larger high temperature phonon peak measured on our floating zone-grown $SmB_6$ single crystals compared to those measured by other groups, indicating very high sample quality of our floating zone-grown single crystals [5]. The considerably larger mean free path of our single crystals compared to samples from refs. [19, 20, 21] explains the boundary-limited phonon behaviour observed in our single crystals of $SmB_6$ which we use to conclude field-induced contribution to thermal conductivity [5]. The difference in mean free path also potentially explains the non-boundary-limited phonon behaviour reported in less pure single crystals of $SmB_6$ [20].

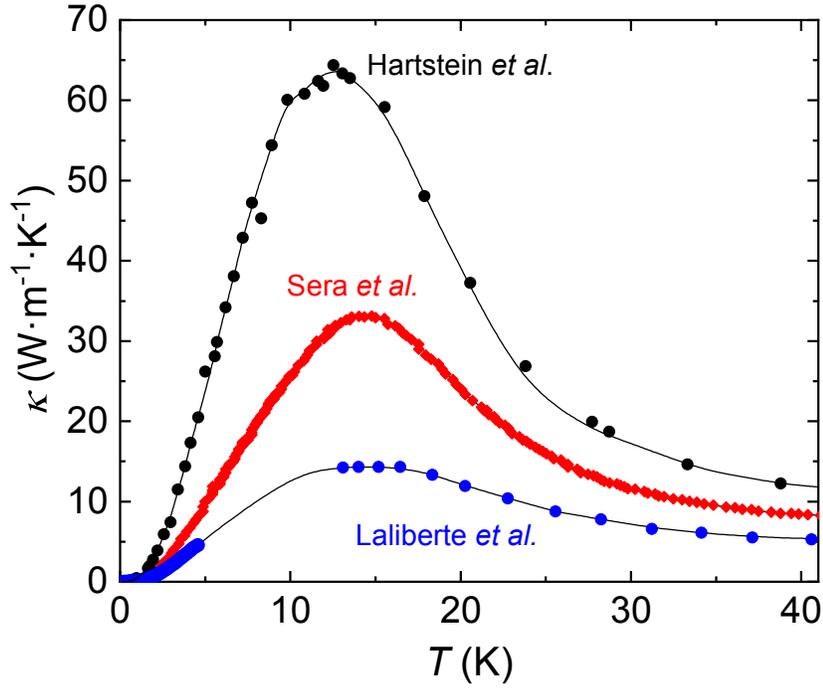

Figure 1: Thermal conductivity $\kappa$ as a function of temperature for a floating zone-grown $SmB_6$ single crystal from the same batch as single crystals on which we measure quantum oscillations (black circles [5, 11]), compared to the thermal conductivity measured for floating zone-grown samples in previous studies (red diamonds [19], blue circles [20]). The size of the peak is found to be largest for the floating zone-grown crystal used in our study, indicating high sample quality.

The size of the low temperature inverse residual resistivity ratio (iRRR) is another indicator of single crystal purity of $SmB_6$ [16, 22, 23]. Extensive screening of over a hundred and thirty samples was performed to select single crystals with the highest values of iRRR for quantum oscillation measurements. Fig. 2(a) compares the resistivity as a function of temperature of the floating zone-grown single crystal $SmB_6$ used for our quantum oscillation measurements, the isotopic floating zone-grown crystal of ref. [12], and crystals grown by various other techniques [24, 25, 26]. We find that single crystals grown by the floating-zone method exhibit the largest values of iRRR. The highest iRRR is found for the floating zone-grown single crystals measured in our quantum oscillation study, which exhibit an iRRR of order $10^5$, exceeding the iRRR in most other single crystals by over an order of magnitude, and with a value of iRRR/thickness ≈ 400, reflecting the highly insulating bulk contribution.

**In-gap low energy excitations in floating zone-grown single crystals of SmB$_6$**

Unconventional low energy excitations within the insulating gap are established by various experimental findings in floating zone-grown single crystals of SmB$_6$ [1, 5]. Fig. 7(c) shows the prominent rise in quantum oscillation amplitude of the 10.6 kT frequency at low temperatures, characteristic of low energy excitations that exhibit a Fermi Dirac-like distribution, further brought out in the inset. Over a broad temperature range, a deviation from the Lifshitz–Kosevich temperature dependence is shown by dashed lines in the main panel of Fig. 7(c). Deviations from the Lifshitz–Kosevich temperature dependence is also observed in other frequencies, as reported in refs. [1, 5]. The low temperature rise in quantum oscillation amplitude contradicts expectations from a gapped model, where the quantum oscillation amplitude would be expected to remain largely flat or to decrease at low temperatures [27].

Furthermore, the linear coefficient of the specific heat capacity has a finite value at low temperatures, providing corroborating evidence for low energy excitations in the gap [5]. Fig. 2(c) shows the measured heat capacity divided by temperature for our floating zone-grown single crystals of SmB$_6$, together with comparative measurements of single crystals presented in ref. [12]. Complementary experiments have shown that an excess value of the linear coefficient of specific heat can arise from magnetic impurities [12, 28]. We find that a finite value of the heat capacity divided by temperature ($\gamma$) is seen for pristine floating zone-grown single crystals used in our quantum oscillation measurements, in which ICP-OES and magnetic characterisation limit any magnetic impurity concentration to within an upper bound of ≈ 0.04%at. Mea-sured finite values of $\gamma \approx 4$ mJ·mol$^{-1}$·K$^{-2}$ (from ref. [5]), and $\gamma \approx 3$ mJ·mol$^{-1}$·K$^{-2}$ (this study), in the vicinity of 2 K are similar to values measured in isotopically pure floating zone-grown single crystals of SmB$_6$ [12, 15], while at least an order of magnitude smaller than the samples impregnated with magnetic impurities [12] and off-stoichiometric samples [16]. The finite low temperature value of the linear coefficient of specific heat, and the low temperature upturn in the linear coefficient of specific heat resembling the low temperature upturn in quantum oscillation amplitude, are hallmarks of low energy excitations in the highest quality floating-zone grown single crystals of insulating SmB$_6$. These features in the linear coefficient of specific heat were also reported in early measurements of isotopically pure floating-zone grown single crystals of SmB$_6$ as identifiers of 'intrinsic coherent state formation within the states at the Fermi energy towards very low temperatures' [15, 29].

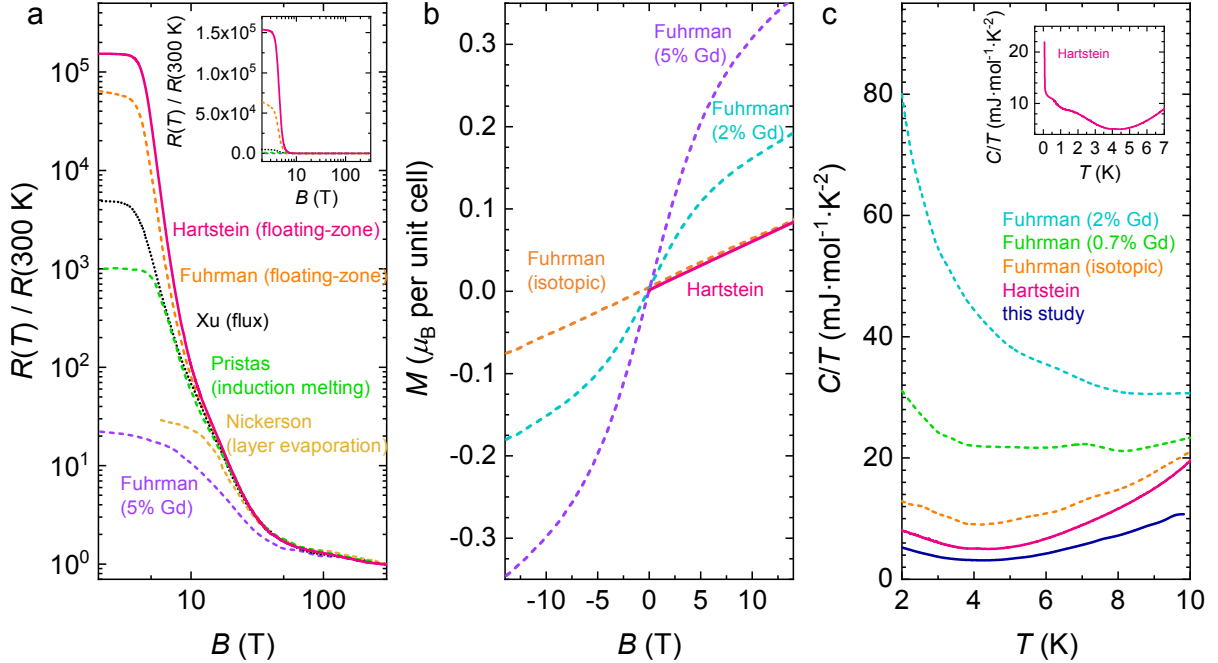

Figure 2: (a) Electrical resistivity as a function of temperature on single crystals of $SmB_6$ grown by different methods, normalised to the resistivity at 300 K. The inverse residual resistivity of floating zone-grown single crystals of $SmB_6$ used to measure quantum oscillations in this study reach a value of the order $10^5$, higher than the isotopic floating zone-grown sample with minimal rare-earth impurities from ref. [12], and more than an order of magnitude higher than samples grown by other growth methods (flux [24], induction melting [25], layer evaporation [26]). The inset shows the same data plotted on a linear scale, further demonstrating the high inverse residual resistivity of our floating zone-grown single crystals. (b) Measured magnetisation of Gd-doped $SmB_6$ samples from ref. [12] (dashed lines in purple and cyan), showing a non-linear magnetisation due to magnetic impurities. Measured magnetisation of floating zone-grown single crystals of $SmB_6$ used for our quantum oscillation studies shows linear paramagnetic behaviour [5], as expected for $SmB_6$ free from magnetic impurities (magenta line), also seen for an isotopic floating zone-grown sample from ref. [12] (orange dashed line). We place an upper bound of $\approx 0.04\%$ at on the magnetic impurity concentration of our sample by fitting with the Langevin function using an effective moment $\mu_{eff} = 7.94\mu_B$ corresponding to the $Gd^{3+}$ state [12]. (c) Measured specific heat capacity of floating zone-grown $SmB_6$ single crystals used in our quantum oscillation measurements (blue and magenta lines [5, 11]), floating zone-grown isotopic sample (orange dashed line [12]) and flux-grown Gd-doped samples (green and cyan dashed lines [12]). The finite linear coefficient of the specific heat capacity is seen to be $\approx 3$ mJ·mol$^{-1}$·K$^{-2}$ in the vicinity of 2 K for our highest purity floating zone-grown single crystals of $SmB_6$ where the magnetic impurity concentration is limited to within $\approx 0.04\%$ at, which is comparable to the finite linear coefficient of the specific heat capacity of isotopically enriched $SmB_6$ also with similarly low magnetic impurity content [15, 28]. The inset shows an upturn in the linear coeffient of the specific heat capacity at low temperatures down to 60 mK as reported in ref. [5], with similarities to the upturn in quantum oscillation amplitude at low temperatures (Fig. 7).

## Intrinsic origin of quantum oscillations from the bulk of floating zone-grown single crystals of SmB$_6$

In this study, as in our previous study [1], quantum oscillation measurements are carried out on floating zone-grown single crystals of SmB$_6$ in the 45 T Hybrid magnet at the National High Magnetic Field Laboratory in Tallahassee using a capacitive torque technique [1]. Complementary techniques have also been used to observe quantum oscillations in these single crystals, including Faraday magnetometry in a superconducting magnet at the Institute for Solid State Physics, University of Tokyo, and magnetic susceptibility in a 65 T pulsed magnet at the National High Magnetic Field Laboratory in Los Alamos [5]. Fig. 3(a-c) shows measured oscillations in the magnetic torque of floating zone-grown SmB$_6$ from the same batch of crystals as the ones employed for the tests of sample quality detailed earlier, and the corresponding oscillation frequencies as found by Fourier transform. Similar to our previous measurements [1], the strongest frequency at low magnetic fields is 500 T, with high frequencies appearing with increasing field; the highest frequency is 15 kT, which appears above 30 T. We find large amplitude quantum oscillations comparable to the paramagnetic torque background (Fig. 4). The large size of quantum oscillations indicates an origin that is not from just a minute fraction of the sample, but rather corresponds to the bulk of the sample.

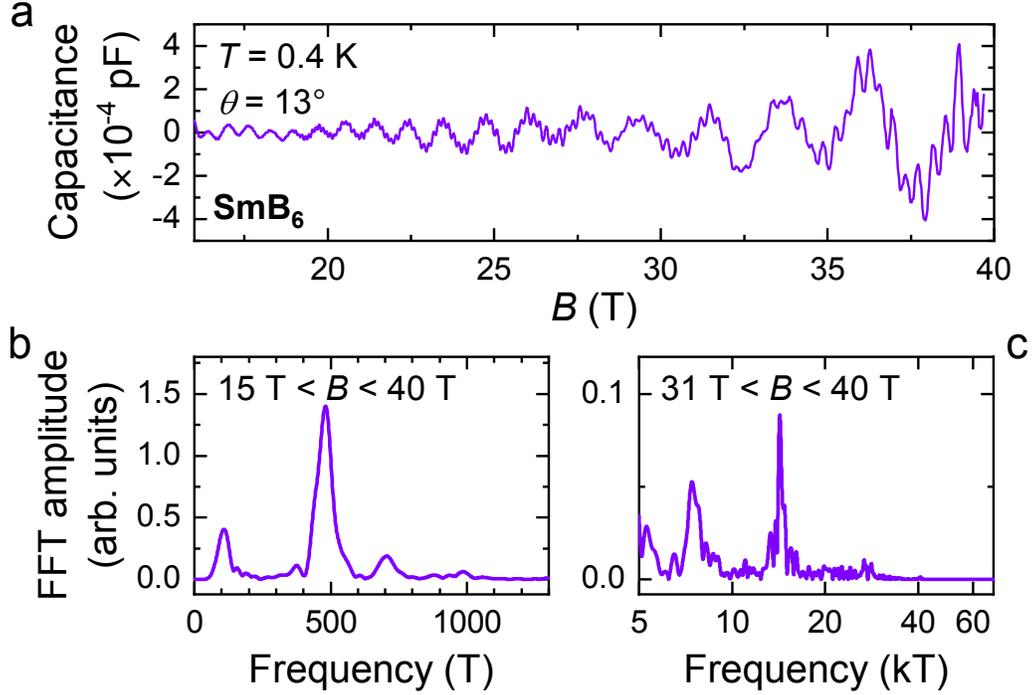

Figure 3: (a) Quantum oscillations in the magnetic torque of floating zone-grown $SmB_6$ single crystals measured in this study as a function of magnetic field. The field is aligned 13° degrees from the [001] direction in the [001]-[111]-[110] rotational plane. (b-c) Fourier transforms of the measured magnetic torque as a function of inverse magnetic field, revealing a wide range of quantum oscillation frequencies ranging from 300 T to 15 kT, similar to our previous study in ref. [1]. A higher magnetic field window is used to reveal the high frequency oscillations corresponding to the $\alpha$, $\lambda$ and $\xi$ branches identified in Fig. 6(a).

We make a quantitative comparison between the absolute size of the measured quantum oscillations and the theoretical amplitude expected for bulk de Haas–van Alphen oscillations within the Lifshitz–Kosevich theory. We convert the measured capacitive torque to absolute units of magnetic moment by using the spring constant of the cantilever, as detailed in ref. [5].

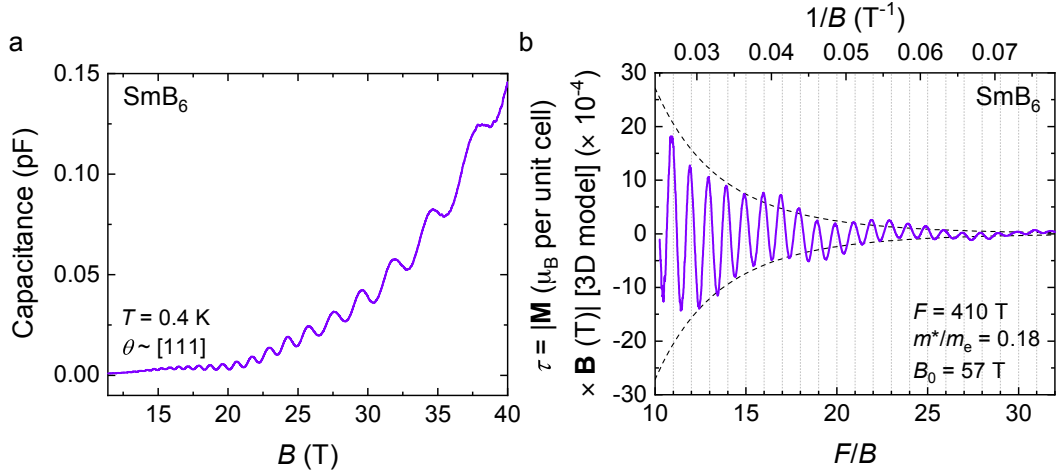

Figure 4: (a) Quantum oscillations in the magnetic torque of a floating zone-grown single crystal of SmB$_6$ are sizeable compared to the paramagnetic background, before any background subtraction. (b) De Haas–van Alphen oscillations in absolute units of magnetic moment corresponding to the $F = 410$ T oscillations after a smooth monotonic background was subtracted. The dashed lines represent the magnetic field dependence of the quantum oscillation amplitude from the impurity scattering (Dingle) damping term for a damping factor $B_0 = 57$ T.

We have cantilever length $L = 3.8$ mm, distance between cantilever and fixed Cu plate $d_0 = 0.1$ mm, spring constant $k = 30$ N·m$^{-1}$, unit cell volume $V_{\text{u.c.}} = a_{\text{u.c.}}^3 = 0.07$ nm$^3$, and crystal volume $s^3 = 0.49$ mm$^3$. We thus convert the measured torque magnetisation in terms of capacitance ($C$) to an absolute magnetic moment $p_s$ in units of Bohr magneton per unit cell by the expression:

$$\Delta p_s = \frac{0.175}{B} \cdot \Delta C \quad \text{T} \cdot \text{pF}^{-1} \; \mu_B \text{ per unit cell.} \qquad (1)$$

The measured quantum oscillatory magnetic moment converted to absolute units for a typical magnetic field sweep is shown in Fig. 4(b). We estimate the theoretical amplitude of the intrinsic quantum oscillatory magnetic moment perpendicular to the applied magnetic field in units of Bohr magneton per unit cell for a three-dimensional Fermi surface using the Lifshitz–Kosevich formula:

$$p_s = D \cdot R_T R_D R_S \cdot \sin(2\pi F/B + \phi) \cdot \sin\theta_M, \qquad (2)$$

where $\sin(2\pi F/B + \phi)$ is the oscillatory term, $\theta_M$ is the angle between the magnetic field $B$ and the total magnetic moment, and $R_T$, $R_D$, and $R_S$ are damping terms due to finite temperature, impurity scattering, and spin-splitting. The exponential damping term $R_D$ is expressed as $R_D = \exp(-B_0/B)$, with damping factor $B_0$ for each frequency. $D$ is the infinite field, zero spin-splitting amplitude given by:

$$D = f(r)\frac{m_\text{e}}{m^*}\left(\frac{a_\text{u.c.}k_\text{F}}{\pi}\right)^3 \sqrt{\frac{B}{8F}}, \tag{3}$$

where $f(r)$ is the anisotropy term, $m^*$ is the effective mass, $k_\text{F}$ is the Fermi wavevector, and $F$ is the oscillation frequency. For the $F = 410$ T frequency oscillations shown in Fig. 4(b), we have $m^*/m_\text{e} = 0.18$, $B_0 = 57$ T, a degeneracy of two and $f(r) = 0.78$ based on the ellipsoidal model we fit with in Fig. 6(a), and estimate $R_S = 0.5\text{-}1$, $\sin\theta_M = 0.1\text{-}1$. The resulting estimate for the theoretical amplitude of quantum oscillations from a three-dimensional Fermi surface is $\approx 6 \times 10^{-4}\text{--}10^{-2}$ $\mu_\text{B}\cdot$ T per unit cell at $B = 30$ T, whilst the measured quantum oscillation amplitude considering a bulk origin is $\approx 10^{-3}$ $\mu_\text{B}\cdot$ T per unit cell, showing consistency between the theoretical estimate for bulk oscillations and the measured size of the oscillations within an order of magnitude. Crucially, the measured quantum oscillation amplitude assuming an origin from only the crystal surface would be at least three orders of magnitude larger than the theoretically expected quantum oscillation amplitude for a two-dimensional Fermi surface (see ref. [5] for a calculation), making this scenario implausible. The same comparison for the high frequencies is hindered by the imprecise estimate of a much higher damping factor due to only high Landau levels being accessed, and requires much higher magnetic fields for a more accurate comparison to be made.

# Comparison of quantum oscillations in floating zone-grown single crystals of $SmB_6$ with quantum oscillations in metallic $LaB_6$

In this study, we perform quantum oscillation measurements on pristine floating-zone grown single crystals of metallic $LaB_6$ to compare with quantum oscillations in $SmB_6$ (Fig. 5(a-c)). We find that quantum oscillations in $SmB_6$ show similar overall behaviour to quantum oscillations in metallic $LaB_6$. A broad range of oscillation frequencies is seen in $LaB_6$, which are observed down to lower fields due to a lower Dingle temperature as compared to $SmB_6$. The size of the observed large quantum oscillations in metallic $LaB_6$ is comparable to expectations for bulk oscillations from the Lifshitz–Kosevich theory [5], as expected for a metal.

The angular dependence of the quantum oscillation frequencies for floating zone-grown $SmB_6$ is shown in Fig. 6(a,b). Measured quantum oscillation frequencies and angular dependence in floating zone-grown $SmB_6$ show similarities with the metallic rare-earth hexaborides [6], which also crystallise in a face-centred cubic crystal structure similar to $SmB_6$. Fig. 6 shows a comparison of the angular dependence of the quantum oscillation frequencies in insulating $SmB_6$ with quantum oscillation frequencies in metallic $LaB_6$. We see a commonality over the entire frequency range with corresponding frequency branches of all major Fermi surface sections. Importantly, we see the high-frequency $\alpha$ branch across the entire angular range for both $SmB_6$ and $LaB_6$, corresponding to the dominant section of the Fermi surface. We also observe the high-frequency $\lambda$ and $\xi$ branches for both materials, which appear for a specific angular range, where two $\alpha$ orbits join through the small neck area [30]. These high-frequency branches are observed across the metallic hexaborides [6], and, as demonstrated below, are not found in Aluminium. The high-frequency branches constitute more than 90% of the Fermi surface sections of the ellipsoidal Fermi surface of metallic hexaborides. As first presented in ref. [1], Fig. 6(a,b) shows the angular dependence of quantum oscillations measured in $SmB_6$ compared with the three-dimensional ellipsoidal Fermi surface model used to model the Fermi surface in the metallic hexaborides [6, 30, 31]. The size of the small connecting ellipsoids, and therefore the $\rho$ frequencies, varies with subtleties of the Fermi surface geometry, and is found to vary between the hexaboride materials.

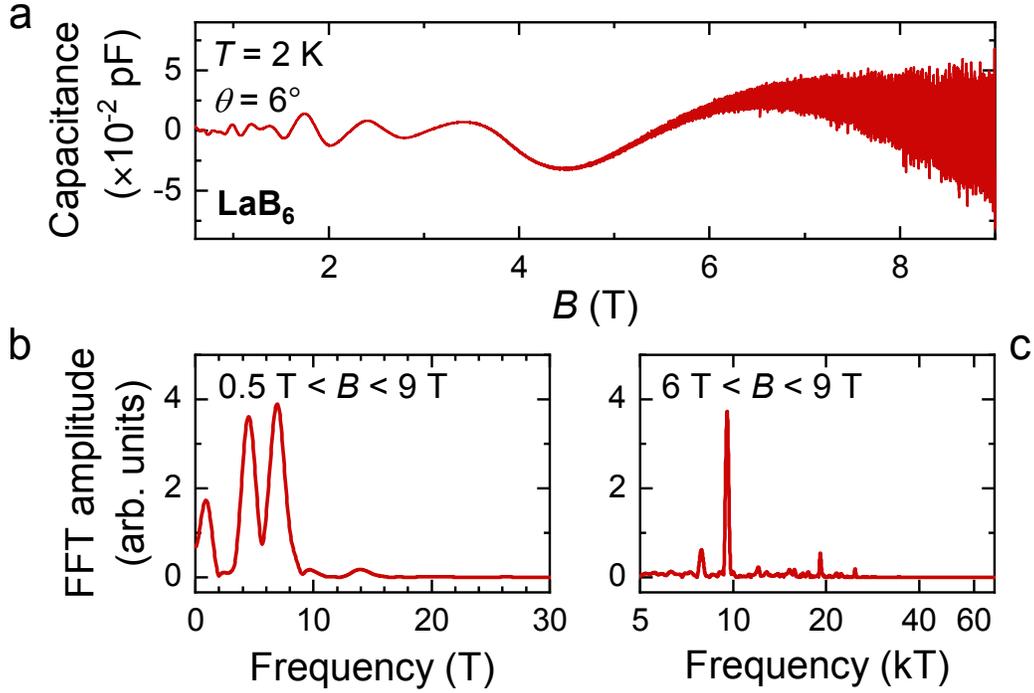

Figure 5: (a) Quantum oscillations in the magnetic torque of floating zone-grown single crystals of LaB$_6$ measured in this study as a function of magnetic field, exhibiting both low- and high-frequency oscillations. The field is aligned 6° degrees from the [001] direction in the [001]-[111]-[110] rotational plane. (b-c) Fourier transforms of the measured magnetic torque as a function of inverse magnetic field, revealing multiple quantum oscillation frequencies up to 10 T corresponding to the $\rho$ branches, and between 8 and 10 kT corresponding to the $\alpha$ branches (see Fig. 6(b)).

Measured quantum oscillation frequencies in SmB$_6$ are characterised by similarly light measured quasiparticle effective masses compared to the conduction electron Fermi surface of LaB$_6$ (Fig. 7). For frequencies originating from the large $\alpha$ ellipsoids, corresponding to the main cross-sections (8 kT $\lessapprox F \lessapprox$ 11 kT), and hole-like orbits that appear for certain angular ranges ($F \gtrapprox$ 800 T), the effective masses of SmB$_6$ scale linearly with the oscillation frequency, very similar to that observed for LaB$_6$. Such a relation between the effective mass and the oscillation frequency is indicative of Fermi surface sections that originate from the same band, a further addition to a suite of evidence for a three-dimensional ellipsoidal Fermi surface from the intrinsic bulk of the material.

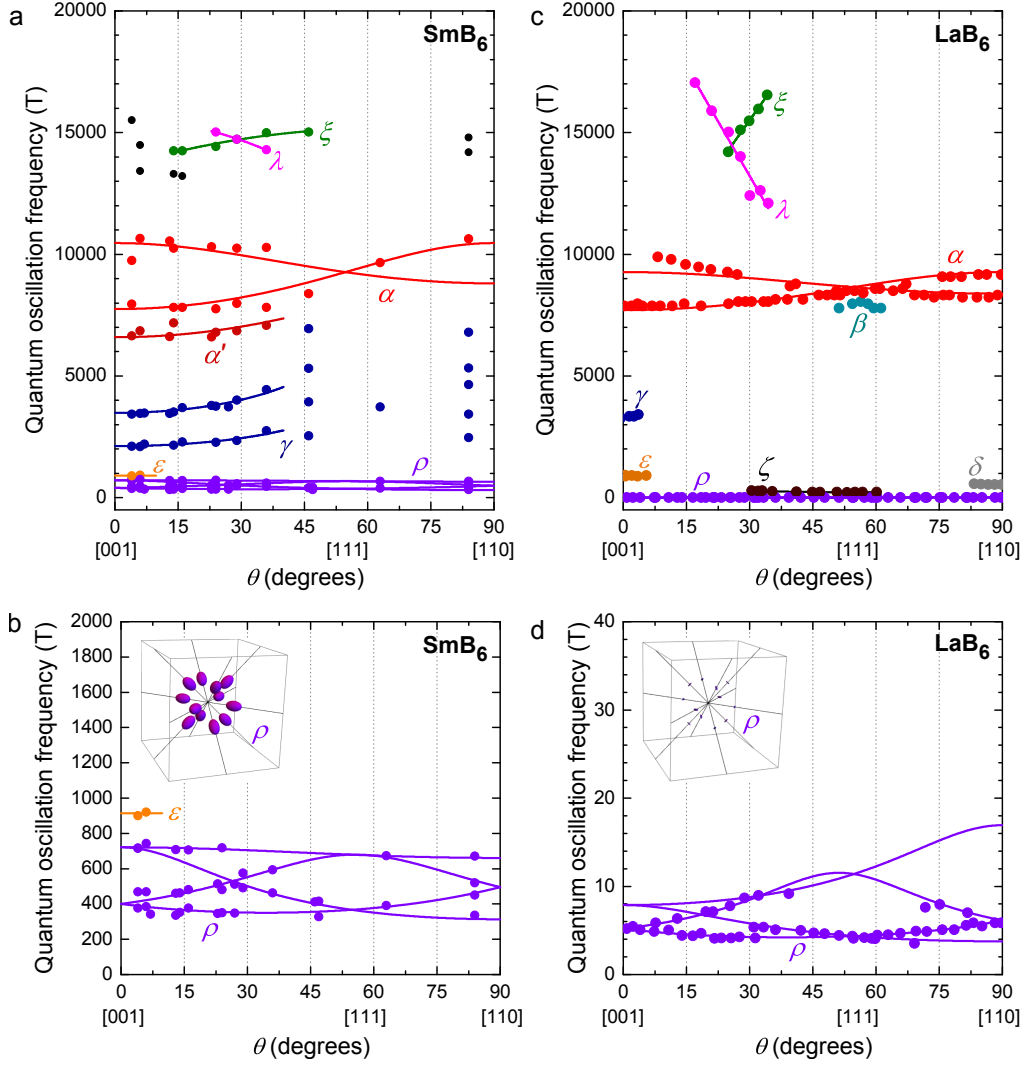

Figure 6: Angular dependence of the quantum oscillation frequencies in the [001]-[111]-[110] rotation plane measured for floating zone-grown single crystals of (a-b) $SmB_6$ from ref. [1], and (c-d) $LaB_6$ from refs. [6, 30], both in good agreement with the angular dependence of a three-dimensional ellipsoidal Fermi surface characteristic of metallic hexaborides. (a) and (c) show the similarity of the higher frequency branches, particularly the main $\alpha$ branches, which correspond to $\approx 50\%$ of the Brillouin zone volume when the three-fold degeneracy is included, while (b) and (d) compare the lower frequency $\rho$ branches, which give rise to the most prominent oscillations, together with illustration of the corresponding Fermi surface sections. Solid lines are based on a three-dimensional ellipsoidal Fermi surface model. For the $\alpha$ branches of $SmB_6$ we obtain a minimum frequency of $F_{min} = 7580$ T, and a ratio of the semi-principal axes of 1.4. For the $\alpha$ branches of $LaB_6$ we obtain $F_{min} = 7710$ T, and a ratio of the semi-principal axes of 1.2. Fitting to the $\rho$ branches of $SmB_6$ we find a minimum frequency of $F_{min} = 313$ T, and the ratios of the two longer semi-principal axes to the shortest axis to be 2.3 and 1.1. In the case of $LaB_6$ we find $F_{min} = 3.7$ T and ratios of 3.5 and 1.5.

## Discussion of quantum oscillations in metallic Aluminium

Recently, a comparison of quantum oscillations in single crystals of Aluminium was made with quantum oscillations in Aluminium flux-grown single crystals of SmB$_6$ [32]. Given that the floating zone-grown single crystals of SmB$_6$ used for our measurements are shown to be free of Aluminium down to the detection limit of $\approx 0.01\%$at by bulk ICP-OES measurements, quantum oscillations in these single crystals are established to be intrinsic to bulk SmB$_6$. These floating zone-grown single crystals of SmB$_6$ thus provide an excellent model system to examine ways in which intrinsic quantum oscillations in SmB$_6$ can be distinguished from quantum oscillations originating from metallic Aluminium. In order to make this comparison, we perform quantum oscillation measurements in pristine single crystals of Aluminium in this study (Fig. 8(a-c)).

Fig. 9 shows the angular dependence of the quantum oscillation frequencies observed for Aluminium. The high and intermediate quantum oscillation frequencies above 2 kT observed in floating zone-grown single crystals of SmB$_6$ are dramatically different from the high quantum oscillation frequencies observed in single crystals of Aluminium. While the low frequency quantum oscillations ($\rho$ frequencies in SmB$_6$ and the $\gamma$ frequencies in Aluminium) have some similar features in the range between 300 T and 500 T, differences become apparent on considering the full angular dependence. The elongated necklace-like Fermi surface of Aluminium yields divergent frequencies along [001] and [111], unlike the three-dimensional ellipsoidal Fermi surface identified in our floating zone-grown single crystals of SmB$_6$, which yields a flat continuum of quantum oscillation frequencies at all angles. In contrast, divergent frequen-cies along several symmetry directions were reported in Aluminium flux-grown single crystals of $SmB_6$ in ref. [33]. A major difference between quantum oscillations observed for floating zone-grown samples of $SmB_6$ and for Aluminium is the appearance of intermediate frequency branches (2–15 kT) in $SmB_6$, corresponding to the main Fermi surface sections, that have no parallels in Aluminium, which is characterized by a non-degenerate 40–80 kT branch at elevated frequencies, which in turn has no parallels in $SmB_6$. Large amplitude oscillations, the intermediate frequency branches (2–15 kT) and the identification of the four flat branches of the $\rho$ frequencies are clear signatures of bulk quantum oscillations in $SmB_6$ corresponding to a three-dimensional Fermi surface.

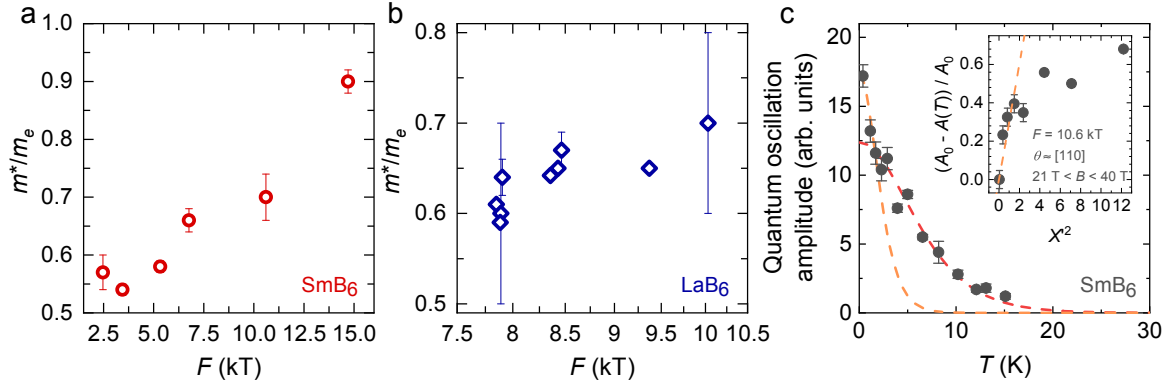

Figure 7: Comparison of the quasiparticle effective masses as a function of oscillation frequency (a) for floating zone-grown $SmB_6$ measured along [110] (for temperatures down to 1 K) from ref. [5], and (b) for $LaB_6$ measured along different field directions based on refs. [6, 30, 34, 35, 36, 37]. The effective mass of frequencies corresponding to the large $\alpha$ ellipsoids in both materials scale roughly linearly with the oscillation frequency, as expected for Fermi surface sections that originate from the same band. (c) The temperature dependence of quantum oscillation amplitude of the 10.6 kT frequency observed in floating zone-grown $SmB_6$ in a magnetic field window of 21 to 40 T and a temperature range 30 K to 550 mK, showing a prominent low-temperature increase. The inset shows the growth of the 10.6 kT frequency amplitude $A(T)$ with respect to $A_0$, the amplitude at the lowest measured temperature, as a function of $X'^2$, where $X' = 2\pi^2 k_B T m_e / e\hbar\mu_0 H_0$, the temperature damping coefficient in the Lifshitz–Kosevich formula, as expected for the Fermi Dirac statistical distribution [38]. The low temperature growth in amplitude is linear in $X'^2$ (shown by orange dashed line), in contrast to the predicted suppression of temperature-dependent amplitude growth at low temperatures for gapped models. The amplitude shows a deviation from the Lifshitz–Kosevich form over the broad temperature range, the red dashed line shows a Lifshitz–Kosevich simulation with effective mass $m^*/m_e = 0.7$ that fits the amplitude as a function of temperature for $T \gtrsim 1$ K, while the orange dashed line shows a Lifshitz–Kosevich simulation with effective mass $m^*/m_e = 1.8$ that fits the amplitude as a function of temperature for $T \lesssim 1$ K. A similar deviation from the Lifshitz-Kosevich temperature dependence is also observed for other frequencies, as reported in refs. [1, 5].

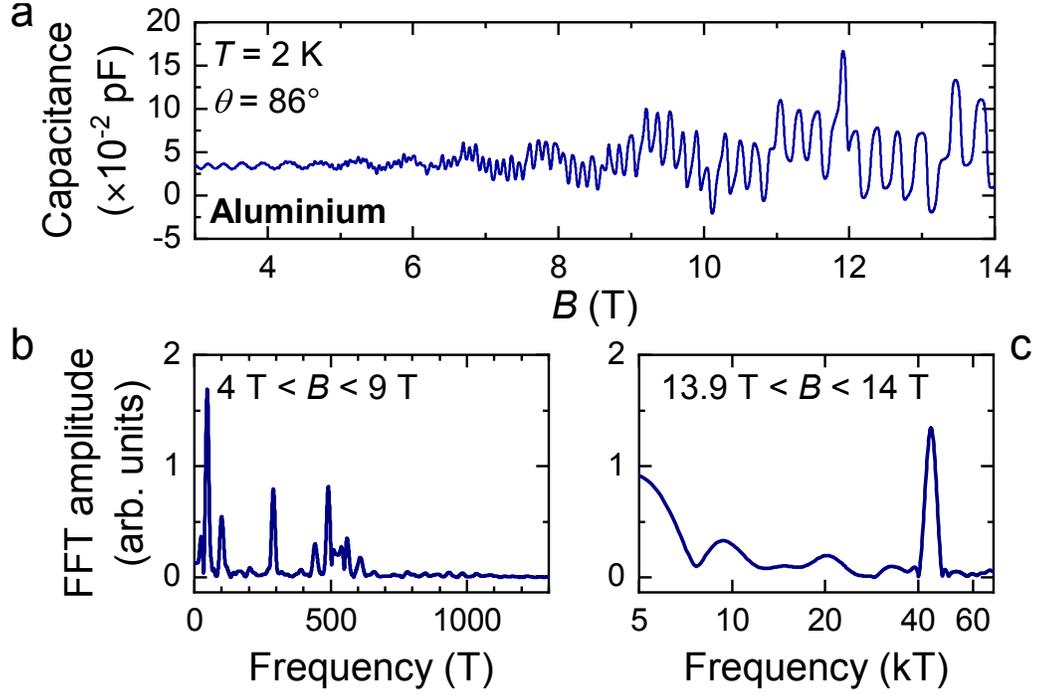

Figure 8: (a) Quantum oscillations in the magnetic torque of Aluminium single crystals measured in this study as a function of magnetic field. The field is aligned 86° degrees from the [001] direction in the [001]-[111]-[110] rotational plane. (b-c) Fourier transforms of the measured magnetic torque as a function of inverse magnetic field, revealing multiple frequency peaks up to 500 T, and a very high frequency peak of 40 kT, when taking a narrow field window, corresponding to the $\psi$ branch (see Fig. 9(b)).

The similarity at the lower end of the frequency branches starting at approximately 300 T between bulk quantum oscillations intrinsic to pristine floating zone-grown $SmB_6$ and quantum oscillations in metallic Aluminium is surprising at first sight. One might consider whether the similarly sized Fermi surfaces are a consequence of the fact that Aluminium and $SmB_6$ both have cubic lattice symmetries, nearly matching lattice constants ($a$ = 4.13 Å in the case of $SmB_6$, and $a$ = 4.05 Å in the case of Aluminium) and the same number of valence electrons. Under these conditions some similarity in the quantum oscillation frequencies might be expected from Luttinger's theorem alone. However, most importantly, the comparison between quantum oscillations intrinsic to floating zone-grown $SmB_6$ and quantum oscillations from metallic Aluminium enables us to clearly outline major differences between these superficially similar quantum oscillations. Such a comparison was not previously possible in ref. [32] or ref. [33] which relied on measurements in Aluminium flux-grown single crystals of $SmB_6$.

## Discussion

We have established intrinsic bulk quantum oscillations in pristine floating zone-grown single crystals of the bulk Kondo insulator $SmB_6$, thus establishing a new class of unconventional insulators, following the first report in ref. [1]. Our findings bring to the fore the question of how such bulk quantum oscillations can arise in a bulk insulator given that thus far they have been considered the preserve of metals. This striking phenomenon first observed in $SmB_6$ [1], now extends across a growing class of unconventional insulators including $YbB_{12}$ [3, 4], and overturns the previously held belief that quantum oscillations are a property unique to metals. A theoretical challenge is thus posed that requires an understanding beyond our current interpretation of quantum oscillations that occur only in metals. The study of more materials in the class of unconventional insulators will provide further clues as to the new theoretical paradigm underlying this rich new phenomenon.

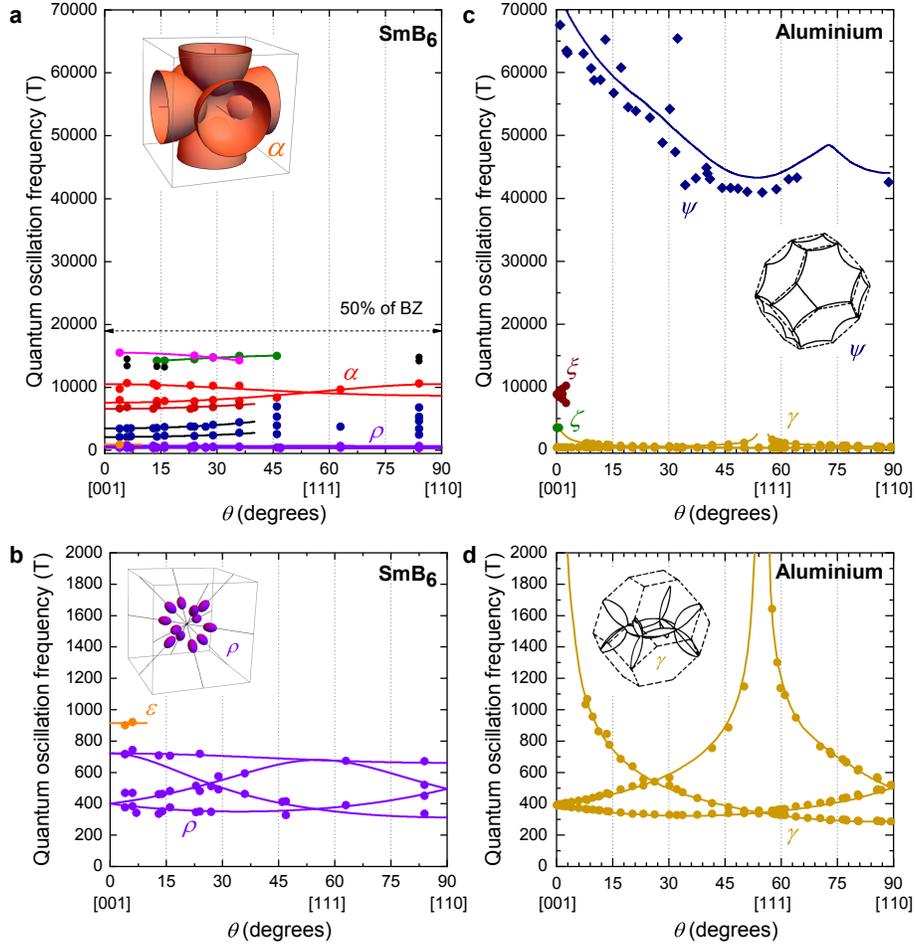

Figure 9: Comparison on the same frequency scale of the measured angular dependence of quantum oscillation frequencies in the [001]-[111]-[110] rotation plane of floating zone-grown $SmB_6$ (a-b) from ref. [1], and Aluminium (c-d) from this study and refs. [39, 40, 41]. The comparison of the high frequency branches in (a) and (c) shows a multitude of branches between 2 kT and 16 kT spanning the entire angular range for $SmB_6$, but none for Aluminium, which has a prominent high-frequency branch above 40 kT (circles are measured by this study, diamonds are from ref. [39], and squares are from refs. [40, 41]). The arrow denotes the maximum expected frequency for half-filling assuming a single spherical Fermi surface section, above which we would not expect to see frequencies for a metallic equivalent, similar to $LaB_6$. (b) $\rho$ branches found for $SmB_6$ have a flat angular dependence corresponding to a Fermi surface of twelve ellipsoids along <110>, as shown in the illustration. (d) Frequencies measured in this study show good overlap with previous measurements (brown circles [40, 41]), and trace the $\gamma$ branches to much higher frequencies than previous measurements, with frequencies as high as 1600 T. A pronounced divergence of the measured frequencies is seen in the [110]-[001] rotation plane, corresponding to a necklace-like Fermi surface of elongated arms as shown in the illustration (taken from ref. [42]). Solid lines follow Ashcroft's model [42] and are from ref. [40]. Frequency points in ref. [41] corresponding to harmonics are not shown. Illustrations of Fermi surface sections corresponding to the $\alpha$ and $\rho$ sections in $SmB_6$ are from ref. [1], while those corresponding to the $\psi$ and $\gamma$ sections in Aluminium are from ref. [42].